\documentclass [noshowpacs,byrevtex,amssymb,superscriptaddress,aps,prl,floatfix,twocolumn]{revtex4}
\usepackage{graphicx}
\usepackage{epsfig}
\usepackage{amsmath}
\usepackage{hyperref}

\usepackage[latin1]{inputenc}
\begin{document}

\title{"Centrifugal fragmentation" in the photodissociation of H$_2^+$ in intense laser fields}

\author{Michael Fischer}
    \affiliation{Institut f{\"u}r Theoretische Physik, Technische
    Universit{\"a}t Dresden, Zellescher Weg 17,
    D--01069 Dresden, Germany} \email[]{E-mail: Michael.Fischer@tu-dresden.de}
\author{Ulf Lorenz}
    \affiliation{HASYLAB at DESY,
    Notkestr. 85,
    D--22607 Hamburg, Germany}
\author{Burkhard Schmidt}
    \affiliation{Institut f{\"u}r Mathematik, Freie Universit{\"a}t Berlin,
    Arnimallee 6, D--14195 Berlin, Germany}
\author{R{\"u}diger Schmidt}
    \affiliation{Institut f{\"u}r Theoretische Physik, Technische
    Universit{\"a}t Dresden, Zellescher Weg 17,
    D--01069 Dresden, Germany} 
\date{\today}

\begin{abstract}
By means of quantum-dynamical and classical trajectory calculations of H$_2^+$ photodissociation in strong laser fields, it is shown that for certain combinations of pulse durations and intensities the rotational dynamics can lead to centrifugal fragmentation. In that case, the photofragments exhibit characteristic angular distributions. The classical calculations provide a transparent physical picture of this mechanism which is also very well established in collisions between atomic nuclei or liquid droplets: non-rotating systems are stable, whereas rotating systems fragment due to the decrease of the fragmentation barrier with increasing angular momentum.
\end{abstract}

\maketitle

Centrifugal fragmentation (CF) is a universal phenomenon of rotating masses. It happens if the angular momentum of the rotating object exceeds a critical value $l_\text{cr}$ and the system breaks into two pieces due to centrifugal forces. In the seminal work of Cohen, Plasil and Swiatecki \cite{Cohen1974557}, these critical $l$-values have been calculated for nuclei, droplets and gravitating masses within the universal rotating liquid drop model (RLDM). Later, critical angular momenta for metallic clusters have been estimated, too \cite{PhysRevA.45.7981}. For nuclei and droplets, the critical angular momenta can be created and measured in heavy ion collisions \cite{bock} and in collisions between macroscopic droplets \cite{brenn1,brenn2,mech}, respectively. During these collisions, an intermediate, fast rotating compound is formed which undergoes fission for $l>l_\text{cr}$ and remains stable for $l<l_\text{cr}$. The $l_\text{cr}$-values are obtained from the universal $\frac{1}{E}$-dependence of the fusion cross sections as function of the incident energy $E$ (see \cite{PhysRevA.45.7981,bock} for details). The experimental  $l_\text{cr}$-values for nuclei and droplets are in excellent agreement with the predictions of the RLDM, although their absolute values differ by more than $20$ orders of magnitude (\cite{PhysRevA.45.7981} and references therein).

Diatomic molecules exposed to strong laser pulses may also rotate because the interaction of the induced or permanent dipole moment of the molecule with the laser field leads to an effective torque towards the laser polarization axis. To what extent rotational effects or even CF may influence the photodissociation process, however, is a challenging, up to date problem \cite{Anis:08a,0953-4075-42-9-091001,PhysRevA.72.045402,njpfischer,tobepub} and the central topic of this work. This will be illustrated for the photodissociation dynamics of nature's most simple molecule, H$_2^+$, where alignment
\cite{Anis:08a,0953-4075-31-13-002,Numico:99a,Atabek:97a} plays an important role for the interpretation of the experimentally observed angular distributions of the fragments (see, e.~g., the summary given in ref. \cite{PhysRevA.72.045402}). 

Molecular alignment is often conveniently discussed in terms of light-dressed, adiabatic Floquet-surfaces \cite{Charron:94a,PhysRevLett.68.3869,Bucksbaum:90a,0953-4075-28-3-006,Shertzer:94a} as a function of the internuclear distance $R$ and the angle $\theta$ between the molecular axis and the laser polarization axis. 
One commonly distinguishes between two different alignment mechanisms: \\
(i) Geometric alignment. 
Upon excitation from the attractive 1s $\sigma_g$ ground state to the repulsive 1s $\sigma_u$ state, a fragmentation channel opens at $\theta=0^{\circ}$ leading to preferential dissociation of molecular ions initially aligned along the polarization axis via the bond softening (BS) mechanism on the lower Floquet surface. In contrast, the upper surface exhibits a barrier at $\theta=0^{\circ}$ (bond hardening (BH) mechanism \cite{Wang94a,PhysRevLett.86.2541,PhysRevLett.68.3869}) and a minimum at $\theta=90^{\circ}$ leading to "counter-intuitive" angular distributions \cite{PhysRevLett.86.2541}. \\
(ii) Dynamic alignment. 
The laser field generates an effective torque towards the laser polarization axis which leads to rotation of initially non-aligned molecules and results in photofragments near $\theta \approx 0^{\circ}$ via BS on the lower Floquet surface \cite{0953-4075-28-3-006,PhysRevA.48.485,PhysRevLett.71.692,PhysRevLett.85.4876,0953-4075-33-14-311,Bucksbaum:90a,PhysRevA.74.043411}. \\
However, experimentally, it is difficult to distinguish between both mechanisms, in particular, because they occur simultaneously with fragments
at $\theta \approx 0^{\circ}$.

In this work, it will be shown that, under certain conditions, the rotational dynamics can lead to an additional photodissociation mechanism. By solving the nuclear time-dependent Schr\"odinger equation (TDSE) a pronounced and characteristic double humped angular distribution of the fragments is found. The first peak centered along the polarization axis results from the well known BS mechanism of geometrically and dynamically aligned molecules. The second peak, however, deviating $15^{\circ}-20^{\circ}$ from the polarization axis, can decisively be attributed to a "centrifugal fragmentation" mechanism where the molecule accumulates sufficient angular momentum that the centrifugal force drives the fragments over the dissociation barrier. 
Under suitable conditions the CF-peak even dominates over the BS-contribution at $\theta = 0^{\circ}$.

The dynamics of H$_2^+$ in an intense laser field $\epsilon(t)$, in
dipole approximation and length gauge, is described by the time-dependent
Schr\"odinger equation (TDSE) for the electron coordinate ($\mathbf{r}$) and the internuclear distance vector ($\mathbf{R}$)
\begin{align} 
\label{TDSEfull}
	\text{i} \frac{\partial}{\partial t} \Psi(\mathbf{R},\mathbf{r},t) 
		&= \left[\hat T_{R} + 
		\hat{H}_e(\mathbf{r};\mathbf{R}) + \hat z \epsilon(t)\right] \Psi(\mathbf{R},\mathbf{r},t) 
\end{align} 
where $\hat T_{R}$ is the kinetic energy of the nuclei, $\hat{H}_e(\mathbf{r};\mathbf{R})$ the electronic Hamiltonian. 
The electric field $\epsilon(t) = \epsilon_0 f(t) \cos{\omega t}$ is linearly polarized along the z-axis and 
characterized by the amplitude $\epsilon_0$, constant carrier frequency $\omega$ and the envelope function
$f(t)$. For this we use a $\sin^2$-shaped turn-on of $15$ fs to simulate a cw-laser and $f(t) = \sin^2{\frac{\pi t}{T}}$ for a pulsed laser with total duration $T$. Atomic units (a.u.) are used unless specified otherwise.

We expand the full wavefunction
$\Psi(\mathbf{R},\mathbf{r},t)$ in parametrically $\mathbf{R}$-dependent eigenstates
$\Phi_{\xi}(\mathbf{r};\mathbf{R})$ of the electronic Hamiltonian as
\begin{align}
\label{BOentw}
	\Psi(\mathbf{R},\mathbf{r},t) &= \sum\limits_{\xi} \frac{1}{R} 
		\Omega_{\xi}(\mathbf{R},t) \Phi_{\xi}(\mathbf{r};\mathbf{R}) \text{\ .}
\end{align}
taking into account the two lowest, i.e., the  $\sigma_g$ and $\sigma_u$ states in the actual calculations, and thus neglect ionization. We have checked that the inclusion of the higher $\pi_u$ state does not change the results for the laser parameters used in this paper.

Within the framework of the Born-Oppenheimer (BO) expansion in eq. (\ref{BOentw}), the nuclear time-dependent
Schr\"odinger equation reads
\begin{align}
	\text{i} \frac{\partial}{\partial t} \Omega_{\xi}(\mathbf{R},t) 
		&= \left[ -\frac{1}{2 \mu} \frac{\partial^2}{\partial R^2} 
		+ \frac{\hat{L}^2}{2 \mu R^2} + V_{\xi}(R) \right] \Omega_{\xi}(\mathbf{R},t)\notag\\
\label{TDSEnuc}
	& + \epsilon(t) \sum\limits_{\eta} D_{\xi\eta}(\mathbf{R}) \Omega_{\eta}(\mathbf{R},t)
\end{align}
with the eigenvalues of the electronic Hamiltonian $V_{\xi}(R)$ and the dipole matrix elements $D_{\xi\eta}(\mathbf{R}) =
\int d^3r \Phi_{\xi}^{*}(\mathbf{r};\mathbf{R}) z \Phi_{\eta}(\mathbf{r};\mathbf{R})$.
The initial state is chosen as a product of the vibrational eigenstate $\nu$, and 
spherical harmonics with fixed angular momentum $l_i$, starting from the
$\sigma_g$-surface
\begin{align}
\label{initstate}
	\Omega_{\sigma_g}(R,\theta,\phi;\nu,l_i,m;t=0) &= \chi_{\nu,l_i}(R) \cdot 
		Y_{l_i}^m (\theta,\phi) \text{\ .}
\end{align}
Equation \eqref{TDSEnuc} is then solved numerically for all $m$-values $-l_i \leq m \leq l_i$ using the code \textsc{WavePacket}
\cite{wavepacket} by expanding the wave function in a basis of plane waves and
spherical harmonics, and applying the split operator scheme \cite{Hermann:88a}.
For the radial grid, an equally-spaced grid with 4096 points ranging from 0.2
a.u. to 236 a.u. is used. Angular momenta up to $l = 59$ are taken into
account. The time step is $10^{-2}$ fs.

\begin{figure}[b]
\parbox{0.47\textwidth}{\includegraphics[width=0.47\textwidth]{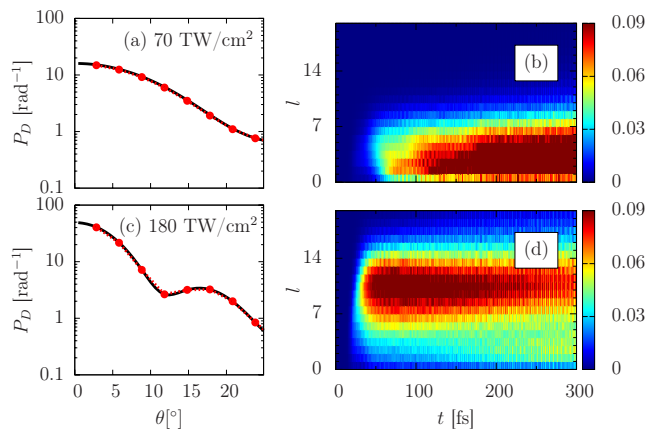}}
\caption{ \label{fig:cw}
	Fragment angular distributions $P_D(\theta)$ (left, solid lines), and corresponding angular momentum
	distributions $P_l(t)$ (right, linear density plots) of dissociating H$_2^+$, initially in its rotational and vibrational ground state, in a 
	cw-laser with $\lambda = 230 \text{\ nm}$, $I = 70 \text{\ TW/cm}^2$ (a,b)  and $I = 180 \text{\ TW/cm}^2$ (c,d).
	The dotted lines in (a), (c) correspond to the reduced one-Floquet-surface calculations (see text).}
\end{figure}
The angular distributions of the photofragments $P_D(\theta)$ are calculated from the nuclear wave
function according to
\begin{align} 
\label{angdis}
	P_D(\theta) &= \frac{1}{2l_i + 1} \sum\limits_{\xi, m} \int\limits_{R_d}^{\infty} dR 
		\int\limits_{0}^{2 \pi} d\phi \Bigl| \Omega_{\xi}(\mathbf{R},T; m) \Bigr|^2		.
\end{align}
with the pulse length T. For the cut-off $R_d$, we have chosen values of 4 a.u. and 10 a.u. for low and high vibrational states, respectively.
They lie beyond the BS dissociation barrier and have been chosen such that further increase does not affect the distribution $P_D$.

The angular momentum distribution of the fragments as a function of time $P_l(t)$, i.e., the weights of the partial waves $l$ contained in the wavepacket, is obtained by
projecting the nuclear wave function $\Omega_{\xi}$ on spherical harmonics $Y_l^m$
\begin{multline}
\label{angmomdis}
	P_l(t) = \frac{1}{2l_i + 1} \int\limits_{R_d}^{\infty} dR \sum\limits_{\xi,m} \\
		\times \biggl| \int\limits_{0}^{\pi} d\theta \sin{\theta} \int\limits_{0}^{2 \pi} 
		d\phi Y_l^{m \,*}(\theta,\phi) \Omega_{\xi}(\mathbf{R},t; m) \biggr|^2 \text{\, .}	
\end{multline}
.

In addition to the BO-expansion (\ref{BOentw}) we solve the TDSE ({\ref{TDSEfull}) by expanding the total wave function $\Psi(\mathbf{R},\mathbf{r},t)$ in time-dependent Floquet states \cite{njpfischer} and consciously take into account only the lowest Floquet surface $V_{\text{F}}(\mathbf{R},t)$  (adiabatic Floquet approximation \cite{Horenko:01a}). The results are compared to those of the full BO-calculations. In doing so we are able to distinguish BS and CF from other dissociation mechanisms (BH, multiphoton dissociation). In this approximation, the nuclear wave function $\Omega_{\text{F}}(\mathbf{R},t)$ is determined by a single TDSE 
\begin{align}
	\text{i} \frac{\partial}{\partial t} \Omega_{\text{F}}(\mathbf{R},t) 
		&= \left[ -\frac{1}{2 \mu} \frac{\partial^2}{\partial R^2} 
		+ \frac{\hat{L}^2}{2 \mu R^2} + V_{\text{F}}(\mathbf{R},t) \right] \Omega_{\text{F}}(\mathbf{R},t) \text{\ .}
\label{TDSEF}	
\end{align}
The angle- and time-dependence of the Floquet surface $V_{\text{F}}(\mathbf{R},t)$ is given implicitly by an effective field strength $\epsilon_{\text{eff}}(\theta,t) = \epsilon_0 f(t) \cos{\theta}$ as $V_{\text{F}}(\mathbf{R},t) = V_{\text{F}}(R,\epsilon_{\text{eff}}(\theta,t))$. The numerical integration of (\ref{TDSEF}) is performed in a similar way as for (\ref{TDSEnuc}).

First, we consider the wavepacket dynamics of H$_2^+$ initially in its ground state ($\nu = 0$, $l_i = 0$),
exposed to a cw laser with a wave length of $\lambda = 230 \text{\ nm}$
and two intensities $I = 70 \text{\ TW/cm}^2$ and $I = 180 \text{\ TW/cm}^2$ (Fig. \ref{fig:cw}). To
ensure the cw-character, i.e., convergence in calculated probabilities, a propagation time of $T = 300$ fs is sufficient.
At $I = 70 \text{\ TW/cm}^2$ a forward peaked angular distribution is obtained  as expected
from the ordinary BS-mechanism on the lowest Floquet-surface with geometric and dynamic alignment contributions.
At $I = 180 \text{\ TW/cm}^2$, however, a second peak at angles between $\theta \approx 15^{\circ} ... \ 20^{\circ} $ occurs.
The time-dependence of the angular momentum distributions in Fig. \ref{fig:cw} (b) and (d) suggests that the appearance of this peak is directly related to distinctly higher $l$-values.
To exclude any other mechanism, caused by multiphoton dissociation or BH
effects, we compare the results of the full BO-calculations with those obtained in the ground-state Floquet approximation (\ref{TDSEF}). 
 Evidently, the results of both calculations are identical and, thus, centrifugal effects are unambiguously the origin of the second maximum in $P_D(\theta)$ ("centrifugal fragmentation").
\begin{figure}[b]
\parbox{0.47\textwidth}{\includegraphics[width=0.47\textwidth]{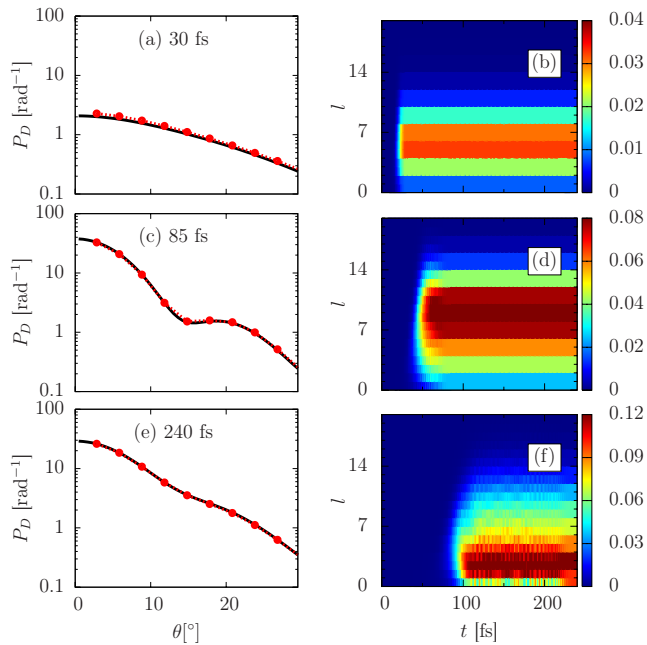}}
\caption{ \label{fig:pulsed} 
	Same as in Fig. \ref{fig:cw}, but for a pulsed laser with $T=$ 30 fs (a,b), 
	85 fs (c,d), 240 fs (e,f) and $\lambda = 230 \text{\ nm}$, $I = 180 \text{\ TW/cm}^2$.}
\end{figure}

Second, we employ a pulsed laser keeping wavelength and intensity unchanged (Fig.~\ref{fig:pulsed}).
For short and long pulses ($T=$ 30 fs and $T=$ 240 fs) single-peaked angular distributions are obtained.
In the fast growing electric field of the short pulse (Fig.~\ref{fig:pulsed} (a)) dissociation of non-aligned molecules becomes also possible resulting in a broad distribution. On the other hand, the slow increase of the field strength in long pulses (Fig.~\ref{fig:pulsed} (e)) supports strongly the dynamical alignment effect which obviously results in a very narrow distribution with a distinct larger maximum value at $\theta = 0^{\circ}$. For intermediate pulse lengths 
($T=$ 85 fs), however, the bimodal distribution is recovered, associated with distinctly larger angular momenta of the fragments as compared to the other cases (Figs.~\ref{fig:pulsed} (b),(d),(f)). The perfect agreement between the exact and restricted one-Floquet surface calculations confirms again the one photon nature of the dynamics in all three cases. Obviously, for the intermediate pulse length the molecule is rotationally accelerated to overcome the dissociation barrier on the first Floquet surface at  $\theta = 0^{\circ}$; the fragments, however, are scattered to final angles $\theta \approx 15 ... 20^{\circ}$ due to their large angular momenta (see also the discussion in terms of classical trajectories, Fig. \ref{fig:floquet}). 
\begin{figure}[b]
\parbox{0.5\textwidth}{\includegraphics[width=0.47\textwidth]{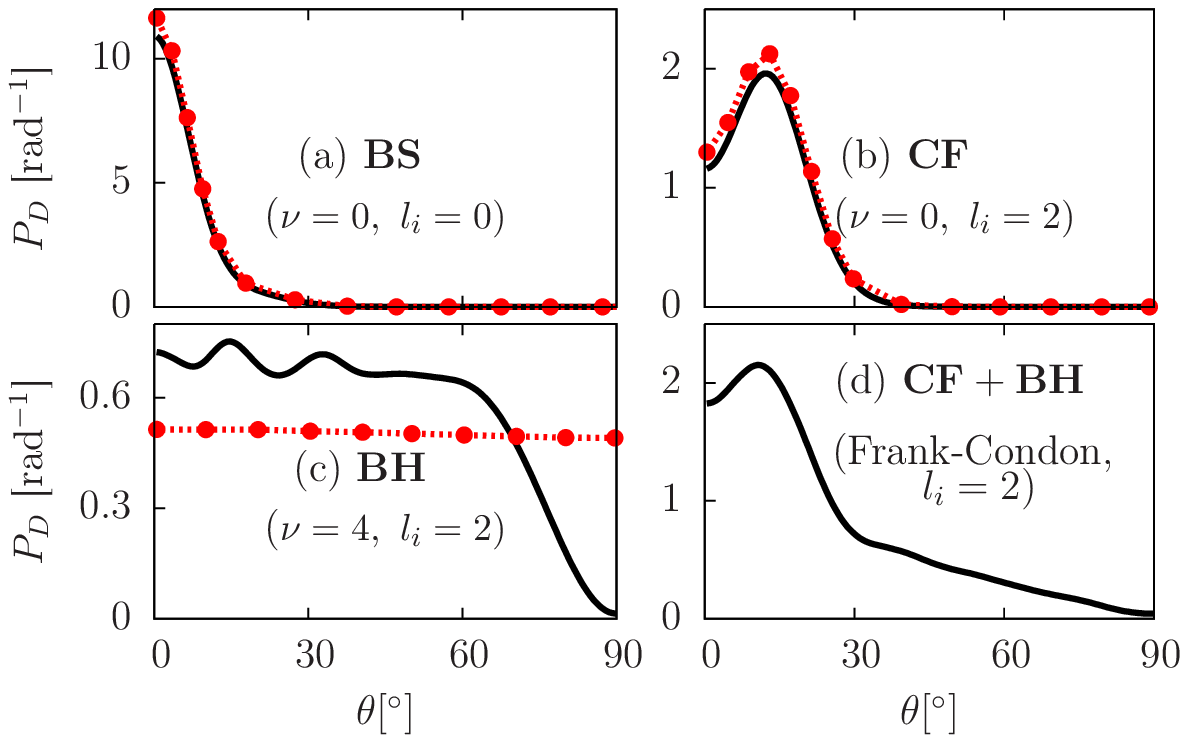}}
\caption{ \label{fig:excited}
	Typical shapes of the angular distributions $P_D(\theta)$ belonging to different dissociation mechanisms (BS, BH and CF)
	emerging in the fragmentation of H$_2^+$ depending on the initial state $\nu$, $l_i$ (solid lines). \\
	The dotted lines correspond to the reduced one-Floquet-surface calculations. The Frank-Condon-averaged distribution
	is shown as well (lower right). The laser parameters are $\lambda = 266 \text{\ nm}$, $T=$ 130 fs and $I = 120 \text{\ TW/cm}^2$.}
\end{figure}

Whereas for H$_2^+$ in its rotational and vibrational ground state ($\nu = 0$, $l_i = 0$) the CF effect shows up as a relatively small contribution in the total angular distribution $P_D(\theta)$ (note the logarithmic scale in Figs.~\ref{fig:cw} and \ref{fig:pulsed}), it can dominate the fragmentation mechanism for state-selected molecules ($l_i > 0$) where $l_i = 2$ is the leading component in the thermal distribution for para-H$_2^+$. This is demonstrated in the upper part of Fig.~\ref{fig:excited} where, for fixed laser parameters, the angular distributions for H$_2^+$ in its rotational and vibrational ground state (Fig.~\ref{fig:excited} (a)) and excited (Fig.~\ref{fig:excited} (b)) H$_2^+$ are compared. The strongly forward peaked distribution for $l_i = 0$ characterizes the ordinary BS mechanism (Fig.~\ref{fig:excited} (a)). Somewhat surprisingly, a minimal shift in the initial angular momentum ($l_i = 2$) leads to a qualitative change in the shape of the distribution with a distinct maximum at finite angles as the fingerprint of the CF mechanism (Fig.~\ref{fig:excited} (b)). It is important to note that the CF maximum in Fig.~\ref{fig:excited} (b) disappears for longer ($300 \text{\ fs}$) and shorter ($30 \text{\ fs}$) pulse durations. This behavior demonstrates again the dynamical nature of the effect (see also Fig.~\ref{fig:pulsed}).

In addition, we consider vibrationally excited molecules (Fig.~\ref{fig:excited} (c),(d)). For the state selected case with $\nu = 4$, $l_i = 2$ a very broad wavelike distribution with an abrupt decrease for angles $\theta \geq 60^{\circ}$ is observed (Fig.~\ref{fig:excited} (c)). The complete disagreement with the isotropic distribution, predicted by the ground-state Floquet calculation (dotted line), proves definitely that this angular distribution is due to the dominating influence of the excited Floquet surface, i.e., the "counterintuitive" BH fragmentation mechanism \cite{PhysRevLett.86.2541}. Hence, the occurrence, and thus the experimental observability of the CF mechanism, depends also on the initial ro-vibrational state of the molecule. 

In many experiments, the neutral H$_2$ molecule serves as the precursor for the fragmenting H$_2^+$ and, thus, a distribution of different $\nu$-values  weighted by their Franck-Condon factors  \cite{FCF} is examined at once. For this case, the predicted angular distribution is shown in (Fig.~\ref{fig:excited} (d)). As can be seen for the actual laser parameters, the CF mechanism "survives" the averaging procedure and leads to an angular distribution with a dominating CF-maximum at finite angles. At the same time, the BH mechanism generates the long tail contribution for angles $\theta \geq 30^{\circ}$ (cf. Fig.~\ref{fig:excited} (b),(d)).
\begin{figure}[t]
\includegraphics[width=0.5\textwidth]{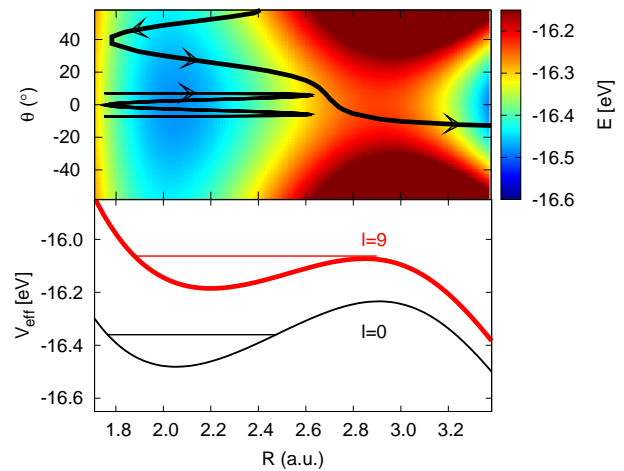}
\caption{ \label{fig:floquet} 
	Classical picture of centrifugal fragmentation.
	Upper: Floquet potential surface $V_F(R,\theta)$ of	H$_2^+$ with two sample trajectories for $\lambda = 230$ nm, $I = 70 \text{\ TW/cm}^2$
	and $T = $ 300 fs (cw).
	Lower: The effective potential $V_{\text{eff}}(R) = V_F(R) + \frac{l^2}{2\mu R^2}$ at $\theta = 0^{\circ}$ for angular momenta $l = 0$ and $l = 9$. The fragmentation barrier is clearly smaller for $l = 9$ due to centrifugal forces. For comparison, the energy of the lowest vibrational state is plotted as straight line.}
\end{figure}

Complementary to the full (\ref{TDSEnuc}) and restricted (\ref{TDSEF}) wavepacket dynamics we have also performed purely classical calculations where the nuclei are propagated on the lowest time-dependent Floquet surface $V_{\text{F}}$ with the Hamiltonian
\begin{align}
	H &=  \frac{P_R^2}{2 \mu} + \frac{L^2}{2 \mu R^2} + V_{\text{F}}(\mathbf{R},t) \text{\ .}
\label{Hamclas}	
\end{align}
The resulting Hamiltonian equations of motion for the momenta ($P_R$,$L$) and coordinates ($R$,$\theta$) represent the classical analogue to Schr\"odinger equation (\ref{TDSEF}). 

Two representative trajectories ($R$,$\theta$) are shown in figure \ref{fig:floquet}.
An initially almost aligned molecule (with initial conditions $\theta_i = 5^{\circ}$, $R_i = 1.8$ a.u.) essentially vibrates	and slightly librates within the potential and remains stable.
A non-aligned molecule with $\theta_i = 60^{\circ}$, $R_i = 2.4$ a.u., however, vibrates and strongly rotates simultaneously, accumulating a large angular momentum and overcoming the barrier at the saddle point near $\theta = 0^{\circ}$, $R \approx 3$ a.u. to finally fragment toward an angle of $\theta \approx -15^{\circ}$.
Also after averaging over trajectory bundles, the CF mechanism is recovered, but overestimated with regard to the quantum calculations due to the lack of tunneling.
However, these calculations provide a clear and simplified physical picture of the underlying mechanism of CF: 
Non-rotating (aligned) molecules are stable, whereas rotating (hence initially non-aligned) molecules fragment owing to the decrease of the fragmentation barrier with increasing angular momentum (compare the effective potentials $V_\text{eff}(R)$ for $l = 0$ and $l = 9$ in the lower part of Fig.~\ref{fig:floquet}).

Exactly the same mechanism leads to the instability (i.e. CF) of fast rotating nuclei, clusters and droplets \cite{Cohen1974557,PhysRevA.45.7981}, where the critical angular momenta are described by the RLDM. In the present molecular case, the $l_\text{cr}$-values depend on the laser parameters via $V_{\text{F}}(R,\theta)$ and their systematic investigation remains an interesting topic for future theoretical studies.

In summary, by means of quantum and classical calculations, we have definitely shown, that the rotational dynamics of H$_2^+$ in intense laser fields can lead to centrifugal fragmentation. The effect shows up in characteristic side-peaked angular distributions of the fragments, provided optimal laser parameters are chosen to accelerate the molecule rotationally above a critical $l$-value. Experimentally, the observability of CF may
be complicated by uncertainties in intensity and focal volume of the laser as well as finite temperature effects, and thus represents a great challenge.


This research is funded by the Danish National Research
Foundation's Center for Molecular Movies and by DFG through program SFB 450.

\end{document}